\documentclass[11pt,a4paper]{article}
\usepackage[hyperref]{acl2021}
\usepackage{times}
\usepackage{latexsym}

\usepackage{bm,amsmath,amssymb,mathrsfs,amsfonts}
\usepackage{caption}
\usepackage{scalefnt}
\usepackage{multirow}
\usepackage{here}
\usepackage{booktabs}
\usepackage{url}
\usepackage{enumitem}
\usepackage{type1cm}
\usepackage{cite}
\usepackage{arydshln}
\usepackage{autobreak}
\usepackage{breqn}
\usepackage{algorithm}
\usepackage{algpseudocode}

\usepackage{hhline} %
\usepackage{highlight}
\renewcommand{\opacity}{50}

\usepackage{pifont}%
\newcommand{\cmark}{\ding{51}}%
\newcommand{\xmark}{\ding{55}}%

\usepackage{microtype}

\aclfinalcopy %

\newcommand{\Sref}[1]{\S\ref{#1}}

\title{ESPnet-ST IWSLT 2021 Offline Speech Translation System}

\author{Hirofumi Inaguma$^{1}$\thanks{*Equal contribution} ~ Brian Yan$^{2*}$ ~ Siddharth Dalmia$^2$ ~ Pengcheng Guo$^3$ \\ ~ \textbf{Jiatong Shi}$^4$ ~ \textbf{Kevin Duh}$^4$ ~ \textbf{Shinji Watanabe}$^{2,4}$ \\
$^1$Kyoto University, Japan ~~ $^2$Carnegie Mellon University, USA ~~ \\ $^3$Northwestern Polytechnical University, China ~~ $^4$Johns Hopkins University, USA\\
\texttt{inaguma@sap.ist.i.kyoto-u.ac.jp}\\ \texttt{byan@cs.cmu.edu}
}

\date{}

\begin{document}
\maketitle
\begin{abstract}
This paper describes the ESPnet-ST group's IWSLT 2021 submission in the offline speech translation track.
This year we made various efforts on training data, architecture, and audio segmentation.
On the data side, we investigated sequence-level knowledge distillation (SeqKD) for end-to-end (E2E) speech translation.
Specifically, we used multi-referenced SeqKD from multiple teachers trained on different amounts of bitext.
On the architecture side, we adopted the Conformer encoder and the Multi-Decoder architecture, which equips dedicated decoders for speech recognition and translation tasks in a unified encoder-decoder model and enables search in both source and target language spaces during inference.
We also significantly improved audio segmentation by using the \texttt{pyannote.audio} toolkit and merging multiple short segments for long context modeling.
Experimental evaluations showed that each of them contributed to large improvements in translation performance.
Our best E2E system combined all the above techniques with model ensembling and achieved 31.4 BLEU on the 2-ref of tst2021 and 21.2 BLEU and 19.3 BLEU on the two single references of tst2021.
\end{abstract}

\section{Introduction}
This paper presents the ESPnet-ST group's English$\to$German speech translation (ST) system submitted to the IWSLT 2021 offline speech translation track.
ESPnet~\citep{watanabe2018espnet} has been widely used for many speech applications; automatic speech recognition (ASR), text-to-speech~\citep{hayashi2020espnet}, speech translation~\citep{inaguma-etal-2020-espnet}, machine translation (MT), and speech separation/enhancement~\citep{li2020espnet}.
The purpose of this submission is not only to show the recent progress on ST researches, but also to encourage future research by building strong systems along with the open-sourced project.

This year we focused on (1) sequence-level knowledge distillation (SeqKD)~\citep{kim-rush-2016-sequence}, (2) Conformer encoder~\citep{gulati2020}, (3) Multi-Decoder architecture~\citep{dalmia2021searchable}, (4) model ensembling, and (5) better segmentation with a neural network-based voice activity (VAD) system~\citep{bredin2020pyannote} and a novel algorithm to merge multiple short segments for long context modeling.
Our primary focus was E2E models, although we also compared them with cascade systems with our best effort.
All experiments were conducted with the ESPnet-ST toolkit~\citep{inaguma-etal-2020-espnet}, and the recipe is publicly available at \url{https://github.com/espnet/espnet/tree/master/egs/iwslt21}.

\section{Data preparation}
In this section, we describe data preparation for each task.
The corpus statistics are listed in Table~\ref{tab:corpus}.
We removed the off-limit talks following previous evaluation campaigns\footnote{\url{https://sites.google.com/view/iwslt-evaluation-2019/speech-translation/off-limit-ted-talks}}.
To fit the GPU memory, we excluded utterances having more than 3000 speech frames or more than 400 characters.
All sentences were tokenized with the {\tt tokenizer.perl} script in the Moses toolkit~\citep{koehn-etal-2007-moses}.

\begin{table}[t]
    \centering
    \small
    \begingroup
    \scalebox{0.95}{
    \begin{tabular}{lcc}\toprule
     & \textbf{\#Hour} & \textbf{\#Sentence}  \\ \midrule
     \textbf{ASR} & &  \\
      \ Must-C & 408 $\times$ 3 & \phantom{0}0.68M \\
      \ Must-C v2 & 458 $\times$ 3 & \phantom{0}0.74M \\
      \ ST-TED (cleaned) & 200 $\times$ 3 & \phantom{0}0.40M \\
      \ Librispeech & 960 & \phantom{0}0.28M \\
      \ TEDLIUM2 & 210 $\times$ 3 & \phantom{0}0.27M \\
      \midrule

     \textbf{E2E-ST} & &  \\
      \ Must-C & 408 $\times$ 3 & \phantom{0}0.68M \\
      \ Must-C v2 & 458 $\times$ 3 & \phantom{0}0.74M \\
      \ ST-TED (cleaned) & 200 $\times$ 3 & \phantom{0}0.40M \\
      \midrule
      
     \textbf{MT} & &  \\
      \ Must-C &  \multirow{9}{*}{-} & \phantom{0}0.68M \\
      \ Must-C v2 &  & \phantom{0}0.74M \\
      \ ST-TED (cleaned) &  & \phantom{0}0.40M \\
      \ Europarl &  & \phantom{0}1.82M \\
      \ Commoncrawl &  & \phantom{0}2.39M \\
      \ Paracrawl &  & 34.37M \\
      \ NewsCommentary &  & \phantom{0}0.37M \\
      \ WikiTitles &  & \phantom{0}1.38M \\
      \ RAPID &  & \phantom{0}1.63M \\
      \ WikiMatrix &  & \phantom{0}1.57M \\ 
      \bottomrule
    \end{tabular}
    }
    \endgroup
    \caption{Corpus statistics}
    \label{tab:corpus}
\end{table}

\subsection{ASR}
We used Must-C~\citep{di-gangi-etal-2019-must}, Must-C v2\footnote{\url{https://ict.fbk.eu/must-c-release-v2-0/}}, ST-TED~\citep{jan2018iwslt}, Librispeech~\citep{librispeech}, and TEDLIUM2~\citep{tedlium} corpora.
We used the cleaned version of ST-TED following~\citep{inaguma19asru}.
The speech data was augmented by three-fold speed perturbation~\citep{speed_perturbation} with speed ratios of 0.9, 1.0, and 1.1 except for Librispeech.
We removed case information and punctuation marks except for apostrophes from the transcripts.
The 5k unit vocabulary was constructed based on the byte pair encoding (BPE) algorithm~\citep{sennrich-etal-2016-neural} with the \texttt{sentencepiece} toolkit\footnote{\url{https://github.com/google/sentencepiece}} using the English transcripts only.

\subsection{E2E-ST}
We used Must-C, Must-C v2, and ST-TED only.
The shared source and target vocabulary of BPE16k units was constructed using cased and punctuated transcripts and translations.

\subsection{MT}
We used available bitext for WMT20\footnote{Europarl, Commoncrawl, Paracrawl, NewsCommentary, WikiTitles, RAPID and WikiMatrix} in addition to the in-domain TED data used for E2E-ST systems.
We first performed perplexity-based filtering with an in-domain n-gram language model (LM)~\citep{moore-lewis-2010-intelligent}.
We controlled the WMT data size by thresholding and obtained three data pools: 5M, 10M, and 20M sentences.
Next, we removed non-printing characters and performed language identification with the \texttt{langid.py} toolkit~\citep{lui-baldwin-2012-langid}\footnote{\url{https://github.com/saffsd/langid.py}} and kept sentences whose language IDs were identified correctly on both English and German sides.
We also removed sentences having more than 250 tokens in either language or a source-target length ratio of more than 1.5 with the \texttt{clean-corpus-n.perl} script in Moses.
Finally, we removed sentences having CJK and other unrelated characters in either language with the built-in \texttt{regex} module in Python.
The resulting data size is shown in Table~\ref{tab:mt_bitext_filtering}.
We found that our filtering strategy removed 22-37\% of data.
Note that the above filtering process was performed over the WMT data only.
For each data size, the joint source and target vocabulary of BPE32k units was constructed using cased and punctuated sentences after the filtering.
We did not use additional monolingual data.

\begin{table}[t]
    \centering
    \small
    \begingroup
    \scalebox{0.95}{
    \begin{tabular}{lccc}\toprule
      \multirow{2}{*}{Filtering method} & \multicolumn{3}{c}{\textbf{\#Sentence}} \\ \cmidrule{2-4}
      & \texttt{WMT5M} & \texttt{WMT10M} & \texttt{WMT20M} \\ \midrule
      In-domain LM & 5.00M & 10.00M & 20.00M \\
      \ + \texttt{langid} & 3.42M & \phantom{0}7.90M & 15.33M \\
      \ \ + length/character & 3.15M & \phantom{0}7.77M & 15.01M \\
      \bottomrule
    \end{tabular}
    }
    \endgroup
    \caption{MT bitext filtering}
    \label{tab:mt_bitext_filtering}
\end{table}

\begin{figure}[t]
    \centering
    \includegraphics[width=0.37\linewidth]{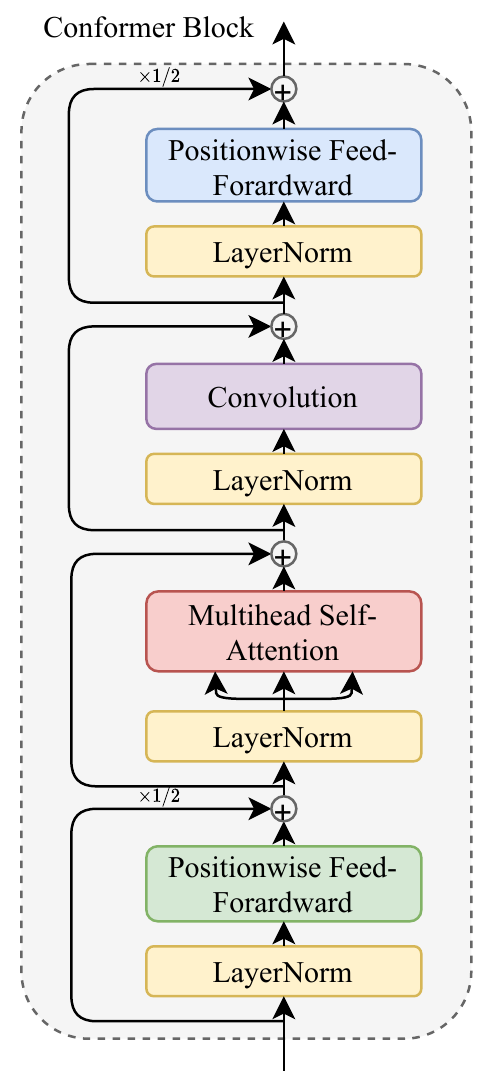}
    \caption{Block diagram of Conformer architecture}
    \label{fig:conformer}
\end{figure}

\section{System}
\subsection{Conformer encoder}
Conformer encoder~\citep{gulati2020} is a stacked multi-block architecture and has shown consistent improvement over a wide range of E2E speech processing applications~\citep{guo2020recent}.
The architecture of each block is depicted in Figure~\ref{fig:conformer}.
It includes a multi-head self-attention module, a convolution module, and a pair of position-wise feed-forward modules in the Macaron-Net style.
While the self-attention module learns the long-range global context, the convolution module aims to model the local feature patterns synchronously.
Recent studies have shown improvements by introducing Conformer in the E2E-ST task~\citep{guo2020recent,inaguma2021source}, which motivated us to adopt this architecture as our system.

\subsection{SeqKD}
Sequence-level knowledge distillation (SeqKD)~\citep{kim-rush-2016-sequence} is an effective method to transfer knowledge in a teacher model to a student model via discrete symbols.
Our recent studies~\citep{inaguma2020orthros,inaguma2021source} showed a large improvement in ST performance with this technique. 
Unlike the previous studies, however, we used more training data than bitext in ST training data to train teacher MT models.
We translated source transcripts in the ST training data by the teacher MT models with a beam width of 5 and then replaced the original ground-truth translation with the generated translation.
We used cased and punctuated transcripts as inputs to the MT teachers.
We also combined both the original and pseudo translations as data augmentation (\textit{multi-referenced training})~\citep{gordon2019explaining}.

\begin{figure}
    \centering
    \includegraphics[width=0.99\linewidth]{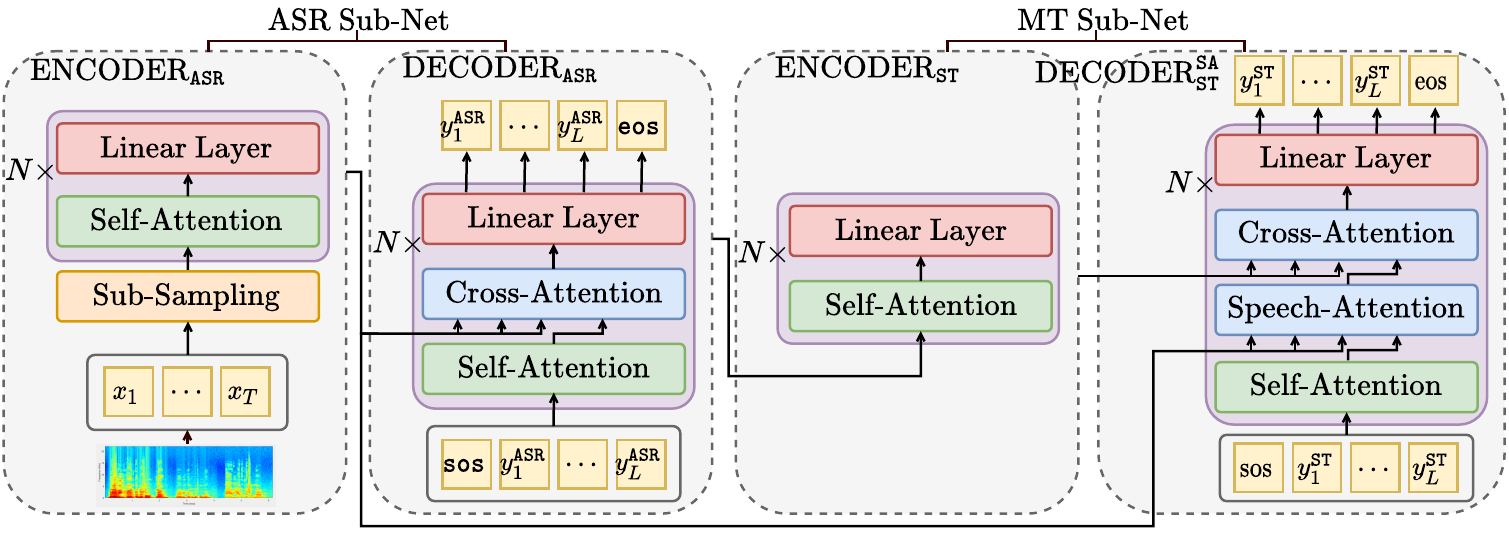}
    \caption{The Multi-Decoder (MD) architecture decomposes the overall ST task with ASR and MT sub-nets while maintaining E2E differentiability.}
    \label{fig:md-fig}
\end{figure}

\subsection{Multi-Decoder architecture}
The Multi-Decoder is an E2E-ST model using Searchable Hidden Intermediates to decompose the overall ST task into ASR and MT sub-tasks~\citep{dalmia2021searchable}.
As shown in Figure~\ref{fig:md-fig}, the Multi-Decoder consists of two encoder-decoder models, an ASR sub-net and a subsequent MT sub-net, where the hidden representations of the ASR decoder are passed as inputs to the encoder of the MT sub-net.
During inference, the best ASR decoder hidden representations are retrieved using beam search decoding at this intermediate stage. 

Since this framework decomposes the overall ST task, it brings several advantages of cascaded approaches into the E2E setting.
For instance, the Multi-Decoder allows for greater search capabilities and separation of speech and text encoding.
However, one trade-off is a greater risk of error propagation from the ASR sub-net to the downstream MT sub-net.
To alleviate this issue, we condition the decoder of the MT sub-net on the ASR encoder hidden representations in addition to the MT encoder hidden representations using multi-source cross-attention.
This improved variant of the architecture is called the Multi-Decoder with Speech Attention.

\subsection{Model ensembling}
\label{sec:ensembling}
We use posterior probability combination to ensemble models trained with different data and architectures.
During inference, we perform a posterior combination at each step of beam search decoding by first computing the softmax normalized posterior probabilities for each model in the ensemble and then taking the mean value.
In this ensembling approach, a single unified beam search operates over the combined posteriors of the models to find the most likely decoded sequence.

\subsection{Segmentation}
How to segment audio during inference significantly impacts ST performances~\citep{Gaido2020,Pham2020,potapczyk-przybysz-2020-srpols,gaido2021beyond}.
This is because the ST systems are usually trained with utterances segmented based on punctuation marks~\citep{di-gangi-etal-2019-must} while the audio segmentation by voice activity detection (VAD) at test time does not access such meta information.
Since VAD splits a long speech recording into chunks by silence regions, it would prevent models from extracting semantically coherent contextual information.
Therefore, it is very important to seek a better segmentation strategy in order to minimize this gap in training and test conditions and evaluate models correctly.
In fact, the last year's winner obtained huge improvements by using their own segmentation strategy.

Motivated by this fact, we investigated two VAD systems apart from the provided segmentation.
Specifically, we used WebRTC\footnote{\url{https://github.com/wiseman/py-webrtcvad}} and pyannote.audio~\citep{bredin2020pyannote}\footnote{\url{https://github.com/pyannote/pyannote-audio}} toolkits.
For WebRTC, we set the frame duration, padding duration, and aggressive mode to 10ms, 150ms, and 3, respectively.
For pyannote.audio, we used a publicly available model pre-trained on the DIHARD corpus~\citep{Ryant2019}.

However, we observed that VAD systems are more likely to generate short segments because they do not take contextual information into account.
Therefore, we propose a novel algorithm to merge multiple short segments into a single chunk to enable long context modeling by self-attention in both encoder and decoder modules.
The proposed algorithm is shown in Algorithm~\ref{algo:merge_segment}.
We first perform VAD and obtain multiple segments.
Then, we check the segments in a greedy way from left to right and merge adjacent segments if (1) the total utterance duration is below a threshold $M_{\rm dur}$ [10ms] and (2) the time interval of the two segments is below a threshold $M_{\rm int}$ [10ms].
This process continues until no segment is merged in an iteration.
Although recent studies proposed similar methods~\citep{potapczyk-przybysz-2020-srpols,gaido2021beyond}, our algorithm is a bottom-up approach while theirs are top-down.

\begin{algorithm}[t]
\caption{Merge short segments after VAD for long context modeling}
\footnotesize
\begin{algorithmic}[1]
\Function{MergeSegment}{${\bm x}, M_{\rm dur}, M_{\rm int}$}
    \State $Q \gets VAD({\bm x})$  {\color{blue} \Comment{$\{(s_{1},e_{1}),\cdots,(s_{M},e_{M})\}$}}
    \While {True}
        \State $N_{\rm merge} \gets 0$
        \State $Q_{\rm next} \gets \{\}$ {\color{blue} \Comment{Queue}}
        \State $S, T \gets s_{1}, e_{1}$ {\color{blue} \Comment{Start/End time}}
        \For {$(s_{m},e_{m}) \in Q$}
            \If {$e_{m} - S < M_{\rm dur}$ and $s_{m} - E < M_{\rm int}$}
                \State $N_{\rm merge} \gets N_{\rm merge} + 1$ {\color{blue} \Comment{Merge segments}}
            \Else
                \State $Q_{\rm next}.enqueue((S,E))$
                \State $S \gets s_{m}$  {\color{blue} \Comment{Reset}}
            \EndIf
            \State $E \gets e_{m}$
        \EndFor
        \State $Q \gets Q_{\rm next}$
        
        \If {$N_{\rm merge} = 0$}
            \State \textbf{break}
        \EndIf
    \EndWhile

    \State \Return $Q$
\EndFunction
\end{algorithmic}\label{algo:merge_segment}
\end{algorithm}

\section{Experimental setting}
In this section, we describe the experimental setting for each task.
The detailed configurations for each task are summarized in Table~\ref{tab:training_config}.

\subsection{Feature extraction}
We extracted 80-channel log-mel filterbank coefficients computed with 25-ms window size and shifted every 10-ms with 3-dimensional pitch features using the Kaldi toolkit~\citep{kaldi}.
The features were normalized by the mean and the standard deviation calculated on the entire training set.
We applied SpecAugment~\citep{specaugment} with mask parameters $(m_{T}, m_{F}, T, F)=(2, 2, 40, 30)$ and time-warping for both ASR and E2E-ST tasks.

\begin{table}[t]
    \centering
    \begingroup
    \scalebox{0.72}{
    \begin{tabular}{lcccc}\toprule
    \multicolumn{1}{c}{\multirow{2}{*}{Configuration}} & \multicolumn{1}{c}{\multirow{2}{*}{ASR}} & \multicolumn{2}{c}{E2E-ST} & \multirow{2}{*}{MT} \\ \cmidrule{3-4}
    \multicolumn{1}{c}{} & \multicolumn{1}{c}{} & non-MD & MD &  \\
    \midrule
     Warmup step & 25k & 25k & 25k & 8k \\
     Learning rate factor & 10.0 & 2.5 & 12.5 & 1.0 \\
     Batch size & 200 utt & 128 utt & 120 utt & 65k tok \\ 
     Epoch & 30 & 30 & 30 & 40 \\
     Validation metric & Accuracy & BLEU & BLEU & BLEU \\
     Model average & 5 & 5 & 5 & 5 \\
     Beam width & 10 & 4 & 16, 10 & 4 \\
     \bottomrule
    \end{tabular}
    }
    \endgroup
    \caption{Summary of training configuration}\label{tab:training_config}
\end{table}

\subsection{ASR}
We used both Transformer and Conformer architectures.
The encoder had two CNN blocks followed by 12 Transformer/Conformer blocks following~\citep{karita2019comparative,guo2020recent}.
Each CNN block consisted of a channel size of 256 and a kernel size of 3 with a stride of 2 $\times$ 2, which resulted in time reduction by a factor of 4.
Both architectures had six Transformer blocks in the decoder.
In both encoder and decoder blocks, the dimensions of the self-attention layer $d_{\rm model}$ and feed-forward network $d_{\rm ff}$ were set to 512 and 2048, respectively.
The number of attention heads $H$ was set to 8.
The kernel size of depthwise separable convolution in Conformer blocks was set to 31.
We optimized the model with the joint CTC/attention objective~\citep{hybrid_ctc_attention} with a CTC weight of 0.3.
We also used CTC scores during decoding but did not use any external LM for simplicity.
We adopted the best model configuration from the Librispeech ASR recipe in ESPnet.

\subsection{MT}
We used the Transformer-Base and -Big configurations in~\citep{vaswani2017attention}.

\subsection{E2E-ST}
We used the same Conformer architecture as ASR except for the vocabulary.
We initialized the encoder parameters with those of the Conformer ASR.
On the decoder side, we initialized parameters like BERT~\citep{devlin-etal-2019-bert}, where weight parameters were sampled from ${\mathcal N}(0, 0.02)$, biases were set to zero, and layer normalization parameters were set to $\beta=0$, $\gamma=1$.
This technique led to better translation performance and faster convergence.

\begin{table}[t]
    \centering
    \begingroup
    \scalebox{0.72}{
    \begin{tabular}{lccc}\toprule
    \multirow{3}{*}{Model} & \multicolumn{3}{c}{\bf{WER} ($\downarrow$)} \\ 
    \cmidrule{2-4}
       & Librispeech & TEDLIUM2 & Must-C \\
       & \texttt{test-other} & \texttt{test} & \texttt{tst-COMMON} \\ 
       \midrule
      Transformer & 9.4 & 6.4 & 7.0	 \\
      Conformer & \bf{7.1} & \bf{6.2} & \bf{5.6} \\
      \bottomrule
    \end{tabular}
    }
    \endgroup
    \caption{Word error rate (WER) of ASR systems}
    \label{tab:result_asr}
\end{table}

\begin{table}[t]
    \centering
    \begingroup
    \scalebox{0.63}{
    \begin{tabular}{lccccccc}\toprule
    \multirow{2}{*}{VAD} & \multirow{2}{*}{$M_{\rm dur}$} & \multirow{2}{*}{$M_{\rm int}$} & \multicolumn{5}{c}{\bf{WER} ($\downarrow$)} \\ 
    \cmidrule(lr){4-7} \cmidrule(lr){8-8}
      &  &  & tst2010 & tst2015 & tst2018 & tst2019 & Avg. \\ \midrule
      \multirow{4}{*}{Provided} 
      & -- & -- & 18.2 & 32.1 & 23.5 & 20.8 & 23.65 \\
      & 1500 & 200 & 14.4 & 29.3 & 18.4 & 15.5 & 19.40 \\
      & 2000 & 200 & 12.7 & 27.7 & 16.4 & 11.5 & 17.08 \\ 
      & 2500 & 200 & 14.5 & 29.9 & 15.1 & 12.2 & 17.93 \\ \midrule

      \multirow{4}{*}{WebRTC} & -- & -- & 35.3 & 35.1 & 44.0 & 22.7 & 34.28 \\
      & 1500 & 200 & 19.4 & 26.7 & 27.7 & 13.8 & 21.90 \\
      & 2000 & 200 & 19.8 & 27.7 & 27.1 & 11.9 & 21.63  \\ 
      & 2500 & 200 & 22.9 & 29.5 & 27.1 & 11.6 & 22.78 \\  \midrule

      \multirow{7}{*}{pyannote} & -- & -- & \phantom{0}9.5 & 24.0 & 15.5 & \phantom{0}7.3 & 14.08 \\ 
      & 1500 & 200 & \phantom{0}8.0 & 23.0 & 12.4 & \phantom{0}7.3 & 12.68 \\
      & 1500 & 100 & \phantom{0}7.5 & 22.2 & 12.4 & \phantom{0}6.5 & 12.15 \\
      & 2000 & 200 & 10.3 & 22.5 & 12.2 & \phantom{0}6.5 & 12.88 \\
      & 2000 & 150 & \phantom{0}9.6 & 21.8 & 12.3 & \phantom{0}6.1 & 12.45 \\
      & 2000 & 100 & \phantom{0}8.1 & \bf{21.5} & \bf{12.0} & \phantom{0}5.8 & 11.90 \\ %
      & 2000 & \phantom{0}50 & \phantom{0}\bf{7.3} & 21.9 & 12.4 & \phantom{0}\bf{5.9} & \bf{11.88} \\
      \bottomrule
    \end{tabular}
    }
    \endgroup
    \caption{Impact of audio segmentation for ASR}
    \label{tab:result_asr_segmenation}
\end{table}

\begin{table*}[t]
    \centering
    \small
    \begingroup
    \scalebox{0.96}{
    \begin{tabular}{lcccccccc}\toprule
      \multirow{3}{*}{Model} & \multicolumn{8}{c}{\bf{BLEU} ($\uparrow$)} \\
     \cmidrule(lr){2-3} \cmidrule(lr){4-4} \cmidrule(lr){5-8} \cmidrule(lr){9-9}
       & \multicolumn{2}{c}{Must-C} & Must-C v2 & \multirow{2}{*}{tst2010} & \multirow{2}{*}{tst2015} & \multirow{2}{*}{tst2018} & \multirow{2}{*}{tst2019} & Must-C \\
       & \texttt{dev} & \texttt{tst-COMMON} & \texttt{tst-COMMON} &  & & & & \texttt{Train} \\
      \midrule
      Base (Must-C only) & -- & 30.02 & 29.86 & 27.28 & 24.92 & 21.13 & 20.37 & \\
      \midrule
      Base (\texttt{WMT5M}) & 31.31 & 34.13 & 33.85 & 31.61 & 32.44 & 28.30 & 28.28 & 45.68 \\
      \ + Big & 27.32 & 29.11 & 28.85 & 27.61 & 28.44 & 24.42 & 23.92 & -- \\
      Base (\texttt{WMT10M}) & \bf{33.28} & \bf{35.09} & 34.80 & \bf{33.58} & 33.26 & 29.24 & 28.87 & 38.31 \\
      \ + In-domain finetune & 30.67 & \bf{35.50} & \bf{35.30} & 30.79 & 31.43 & 25.35 & 26.10 & -- \\ 
      Base (\texttt{WMT20M}) & 33.15 & 35.06 & \bf{34.87} & 33.26 & \bf{33.56} & \bf{29.94} & \bf{29.08} & 33.60 \\
      
      \bottomrule
    \end{tabular}
    }
    \endgroup
    \caption{BLEU scores of text-based MT systems}
    \label{tab:result_mt}
\end{table*}

\section{Results}
\subsection{ASR}
\subsubsection{Architecture}
We compared Transformer and Conformer ASR architectures in Table~\ref{tab:result_asr}.
We observed that Conformer significantly outperformed Transformer.
Therefore, we use the Conformer encoder in the following experiments.

\subsubsection{Segmentation}\label{sssec:segmentation_asr}
Next, we investigated the VAD systems and the proposed segment merging algorithm for long context modeling in Table~\ref{tab:result_asr_segmenation}.
We used the same decoding hyperparameters tuned on Must-C.
We firstly observed that merging short segments was very effective probably because it alleviated frame classification errors in the VAD systems.
Among three audio segmentation methods, we confirmed that pyannote.audio significantly reduced the WER while WebRTC had negative impacts compared to the provided segmentation.
Specifically, we found that the \texttt{dihard} option in pyannote.audio worked very well while the rest options did not.
The optimal maximum duration $M_{\rm dur}$ was around 2000 frames (i.e., 20 seconds).
In the last experiments, we tuned the maximum interval $M_{\rm int}$ among \{50, 100, 150, 200\} and found 50 and 100 (i.e., 0.5 and 1 second) was best on average.
Compared to the provided segmentation, we obtained a 49.6\% improvement on average.

\begin{table*}[t]
    \centering
    \begingroup
    \scalebox{0.66}{
    \begin{tabular}{llccccccccc}\toprule
    \multirow{3}{*}{ID} & \multirow{3}{*}{Model} & \multicolumn{8}{c}{\bf{BLEU} ($\uparrow$)} \\ 
    \cmidrule(lr){3-5} \cmidrule(lr){6-6} \cmidrule(lr){7-10}
       & & \multicolumn{3}{c}{Must-C} & Must-C v2 & \multirow{2}{*}{tst2010} & \multirow{2}{*}{tst2015} & \multirow{2}{*}{tst2018} & \multirow{2}{*}{tst2019} \\
       & & \texttt{dev} & \texttt{tst-COMMON} & \texttt{tst-HE} & \texttt{tst-COMMON} \\ 
       \midrule

       \multirow{6}{*}{-} & Bidir SeqKD (E2E)~\citep{inaguma2021source} & 25.67 & 27.01 & 25.36 & -- & -- & -- & -- & -- \\
       & Multi-Decoder (E2E)~\citep{dalmia2021searchable} & -- & 26.4\phantom{0} & -- & -- & -- & -- & -- & -- \\
       & RWTH (Cascade)~\citep{bahar2021tight} & -- & 26.50 & 26.80 & -- & -- & 28.4\phantom{0} & -- & -- \\
       & KIT (E2E)~\citep{Pham2020} & -- & 30.60 & -- & -- & 24.27 & 21.82 & -- & -- \\
       & KIT (Cascade)~\citep{Pham2020} & -- & -- & -- & -- & 26.68 & 24.95 & -- & -- \\
       & SRPOL (E2E)~\citep{potapczyk-przybysz-2020-srpols} & -- & -- & -- & -- & 29.44 & 24.6\phantom{0} & -- & 23.96 \\
      \midrule
      
      \texttt{A1} & Baseline (\texttt{X}) & 25.14 & \bf{35.63} & 22.63 & \bf{36.07} & 21.40 & 18.18 & 16.69 & 17.39 \\
      \texttt{A2} & \ + SeqKD (\texttt{Y}) & 26.31 & 29.29 & 26.33 & 29.50 & 23.34 & 21.24 & 21.09 & 22.25 \\
      \texttt{A3} & \ + 2ref SeqKD (\texttt{X}+\texttt{Y}) & 26.50 & 30.59 & 26.21 & 30.92 & 23.00 & 22.18 & 20.38 & 21.59 \\
      \texttt{A4} & \ + 3ref SeqKD (\texttt{X}+\texttt{Y}+\texttt{Z}) & \bf{27.66} & 30.90 & \bf{27.44} & 31.07 & \bf{24.97} & \bf{22.66} & \bf{22.20} & \bf{23.41} \\ \midrule
    
      \texttt{B1} & MD + 2ref SeqKD & -- & 30.78 & -- & -- & -- & -- & -- & \textbf{23.78} \\ \midrule
      
      \texttt{C1} & Conformer ASR $\to$ Base MT (\texttt{WMT10M}) & 27.01 & 29.42 & 26.13 & 29.75 & \bf{25.04} & \bf{23.17} & \bf{23.05} & 23.19 \\
      
      \bottomrule
    \end{tabular}
    }
    \endgroup
    \caption{BLEU scores of ST systems. \texttt{X}: original, \texttt{Y}: \texttt{WMT5M}, \texttt{Z}: \texttt{WMT10M}. For unsegmented test sets, we used \underline{pyannote.audio} with $M_{\rm dur}=2000$ and $M_{\rm int}=100$.}
    \label{tab:result_st}
\end{table*}

\subsection{MT}
In this section, we show the results of our MT systems used for cascade systems and pseudo labeling in SeqKD.
We report case-sensitive detokenized BLEU scores~\citep{papineni-etal-2002-bleu} with the \texttt{multi-bleu-detok.perl} script in Moses.
We carefully investigated the effective amount of WMT training data to improve the performance of the TED domain.
The results are shown in Table~\ref{tab:result_mt}.
We confirmed that adding the WMT data improved the performance by more than 4 BLEU.
Regarding the WMT data size, using up to 10M sentences was helpful, but 20M did not show clear improvements, probably because of the undersampling of the TED data. 
Oversampling as in multilingual NMT~\citep{arivazhagan2019massively} could alleviate this problem, but this is beyond our scope.

After training with a mix of the WMT and TED data, we also tried to finetune the model with the TED data only, but this did not lead to clear improvement, especially for the IWSLT test sets.
Increasing the model capacity was not helpful, although the conclusion might change by adding more training data and evaluating the model in other domains such as news.
Because our primary focus to use MT systems was pseudo labeling for SeqKD, we decided to use the Base configuration to speed up decoding.

Finally, we checked the BLEU scores on the Must-C training data used for SeqKD.
We observed that adding more WMT data decreased the BLEU score, from which we can conclude that using more WMT data gradually changed the MT output from the TED style.
Therefore, we decided to use the models trained on \texttt{WMT5M} and \texttt{WMT10M} as teachers for SeqKD.

\subsection{Speech translation}

\subsubsection{E2E-ST}

\paragraph{SeqKD} The results are shown in Table~\ref{tab:result_st}.
We first observed the baseline Conformer model (\texttt{A1}) achieved 35.63 BLEU on the Must-C \texttt{tst-COMMON} set, and it is the new state-of-the-art record to the best of our knowledge. 
Surprisingly, it even outperformed text-based MT systems in Table~\ref{tab:result_mt}.
On the other hand, unlike our observations in~\citep{inaguma2020orthros,inaguma2021source}, SeqKD (\texttt{A2-4}) degraded the performance on the Must-C \texttt{tst-COMMON} set.
However, the results on the Must-C \texttt{dev} and \texttt{tst-HE} sets showed completely different trends, where we observed better BLEU scores by SeqKD in proportion to the WMT data used for training the teachers.
Therefore, after tuning audio segmentation, we also evaluated the models on the unsegmented IWSLT test sets.
Here, we used the pyannote.audio based segmentation with $(M_{\rm dur},M_{\rm int})=(2000,100)$ as described in~\Sref{sssec:segmentation_asr}.
Then, we confirmed large improvements with SeqKD by 2-6 BLEU, and therefore we decided to determine the best model based on the IWSLT test sets.
Multi-referenced training consistently improved the BLEU scores on the IWSLT sets.
For example, \texttt{A4} outperformed \texttt{A1} by 6.02 BLEU on tst2019 although the tst2019 set was well-segmented (WER: 6.0\%).
Given these observations, we recommend evaluating ST models on multiple test sets for future research.

\paragraph{Multi-Decoder architecture}
We combined the SeqKD and Multi-Decoder techniques in our \texttt{B1} system.
\texttt{B1}, which used a conformer ASR encoder and \texttt{2ref} SeqKD, showed an improvement of 2.19 BLEU on tst2019 over \texttt{A3}, the encoder-decoder which also used \texttt{2ref} SeqKD.
\texttt{B1} also achieved a slightly higher result on tst2019 compared to \texttt{A4} which used \texttt{3ref} SeqKD.
These results suggest that the Multi-Decoder architecture is indeed compatible with SeqKD.

\begin{table}[t]
    \centering
    \small
    \begin{tabular}{llc}\toprule
      ID & Ensembled Models & tst2019 \\  
      \midrule
      \phantom{0}- & \texttt{B1} & 21.06 \\
      \texttt{E1} & \texttt{B1}, \texttt{A4} & 22.51 \\
      \texttt{E2} & \texttt{B1}, \texttt{A4}, \texttt{A1} & 22.83 \\
      \texttt{E3} & \texttt{B1}, \texttt{A4}, \texttt{A1}, \texttt{A3} & 23.36 \\
      \texttt{E4} & \texttt{B1}, \texttt{A4}, \texttt{A1}, \texttt{A3}, \texttt{A2} & \textbf{23.61} \\
      \bottomrule
    \end{tabular}
    \caption{BLEU ($\uparrow$) scores of ensembled E2E-ST systems on tst2019, using the \underline{provided} segmentation with $M_{\rm dur}=2000$ and $M_{\rm int}=100$}
    \label{tab:result_ensembling}
\end{table}

\begin{table}[t]
    \centering
    \begingroup
    \scalebox{0.63}{
    \begin{tabular}{lccccccc}\toprule
    \multirow{2}{*}{VAD} & \multirow{2}{*}{$M_{\rm dur}$} & \multirow{2}{*}{$M_{\rm int}$} & \multicolumn{5}{c}{\bf{BLEU} ($\uparrow$)} \\ 
    \cmidrule(lr){4-7} \cmidrule(lr){8-8}
       &  &  & tst2010 & tst2015 & tst2018 & tst2019 & Avg. \\ \midrule
       Provided${}^\dagger$ & -- & -- & -- & -- & -- & 20.1\phantom{0} & -- \\ \midrule

      \multirow{4}{*}{\shortstack{Provided\\(E2E)}} 
      & -- & -- & 21.99 & 19.94 & 19.29 & 19.70 & 20.23 \\
       & 1000 & 200 & 22.62 & 20.54 & 19.80 & 20.54 & 20.88 \\
       & 1500 & 200 & 23.00 & 21.66 & 20.14 & 21.50 & 21.58 \\
       & 2000 & 200 & 22.95 & 21.58 & 20.03 & 21.34 & 21.48 \\
      \midrule
    
       \multirow{4}{*}{\shortstack{WebRTC\\(E2E)}} 
       & -- & -- & 13.13 & 12.97 & 11.07 & 13.32 & 12.62 \\
       & 1000 & 200 & 20.95 & 20.66 & 17.09 & 20.87 & 19.89 \\
       & 1500 & 200 & 21.00 & 20.99 & 17.67 & 21.05 & 20.18 \\
       & 2000 & 200 & 20.25 & 21.81 & 17.08 & 20.71 & 19.96 \\
       \midrule
       
       \multirow{7}{*}{\shortstack{pyannote\\(E2E)}}
       & -- & -- & 22.26 & 16.84 & 17.78 & 19.98 & 19.22 \\ 
       & 1500 & 200 & 25.00 & 22.22 & 21.97 & 22.67 & 22.97 \\
       & 1500 & 100 & \bf{25.92} & \bf{22.81} & \bf{22.51} & 22.88 & \bf{23.53} \\
       & 2000 & 200 & 24.10 & 21.98 & 21.00 & 22.71 & 22.45 \\
       & 2000 & 150 & 24.25 & 22.26 & 21.41 & 22.99 & 22.73 \\
       & 2000 & 100 & 24.97 & 22.66 & 22.20 & \bf{23.41} & 23.31 \\
       & 2000 & \phantom{0}50 & 24.50 & 20.67 & 22.14 & 22.89 & 22.55 \\ \midrule \midrule
       
       \multirow{6}{*}{\shortstack{pyannote\\(Cascade)}} 
       & 1500 & 200 & 25.06 & 22.65 & 23.01 & 22.51 & 23.31 \\
       & 1500 & 100 & \bf{25.56} & 22.85 & 23.03 & 22.82 & 23.57 \\
       & 2000 & 200 & 24.41 & 22.76 & 22.15 & 22.08 & 22.85 \\
       & 2000 & 150 & 24.50 & 23.03 & \bf{23.12} & 23.11 & 23.44 \\
       & 2000 & 100 & 25.04 & \bf{23.17} & 23.05 & \bf{23.19} & \bf{23.61} \\
       & 2000 & \phantom{0}50 & 24.33 & 20.79 & \bf{23.12} & 23.11 & 22.84 \\

      \bottomrule
    \end{tabular}
    }
    \endgroup
    \caption{Impact of audio segmentation for ST. \texttt{A4} was used for the E2E model. ${}^\dagger$~\citep{potapczyk-przybysz-2020-srpols}}
    \label{tab:result_st_segmenation}
\end{table}

\paragraph{Model ensemble}
As shown in Table~\ref{tab:result_ensembling}, ensembling our various ST systems using the posterior combination method described in~\Sref{sec:ensembling} showed improvements over the best single model, \texttt{B1}.
We found that an ensemble of all of our models, \texttt{A1-4} and \texttt{B1}, achieved the best result of 23.61 BLEU on tst2019 and outperformed \texttt{B1} by 2.55 BLEU.
Although \texttt{A1} as a single system performs worse on tst2019 than the other single systems as shown in Table~\ref{tab:result_st}, including it in an ensemble with the two best single systems, \texttt{B1} and \texttt{A4}, still yielded a slight gain of 0.32 BLEU (\texttt{E2}).
Therefore, we can conclude that weak models are still beneficial for ensembling.

\begin{table*}[t]
    \centering
    \small
    \begin{tabular}{lcccccccc}\toprule
      \multirow{3}{*}{System} & \multirow{3}{*}{Segmentation} & \multirow{3}{*}{\shortstack{Segment\\merging}} & \multirow{3}{*}{$M_{\rm int}$} &  \multicolumn{5}{c}{\bf{BLEU} ($\uparrow$)} \\ 
    \cmidrule(lr){5-6} \cmidrule(lr){7-9}
      &  &  & & \multirow{1}{*}{tst2019} & \multirow{1}{*}{tst2020} & \multicolumn{3}{c}{tst2021} \\
       &  & & & & & ref1 & ref2 & both \\  
      \midrule
      \multirow{2}{*}{IWSLT'20 winner${}^{\clubsuit}$} & given & -- & -- & 20.1\phantom{0} & 21.5 & -- & -- & -- \\
      & own & -- & -- & 23.96 & 25.3 & -- & -- & -- \\
      \midrule
      
      \texttt{E4} (primary) & pyannote & \cmark & 200 & \bf{24.14} & \bf{25.6} & \bf{19.3} & \bf{21.2} & \bf{31.4} \\ %
      \midrule
      \texttt{E4}+* & pyannote & \cmark & 200 & 24.41 & 25.5 & \bf{19.7} & 20.6 & 30.8 \\  %
      \texttt{E4}+* & pyannote & \cmark & 100 & \bf{24.87} & \bf{26.0} & 19.5 & 21.1 & 31.3 \\  %
      
      \texttt{E4}+* & given & \cmark & 100 & 23.72 & 25.1 & 19.4 & \bf{21.4} & \bf{31.5} \\ %
      \texttt{E4}+* & given & \xmark & -- & 21.10 & 22.3 & 17.4 & 18.4 & 27.7 \\ %
      \texttt{B1} & pyannote & \cmark & 100 & 23.78 & 25.0 & 18.9 & 20.9 & 31.1 \\ %
      
      \bottomrule
    \end{tabular}
    \caption{BLEU scores of submitted systems on tst2020 and tst2021. ${}^{\clubsuit}$~\citep{potapczyk-przybysz-2020-srpols}. $M_{\rm dur} =2000$ was used for the segment merging algorithm. *Late submission (not official). \texttt{E4}+ denotes \texttt{E4} trained for more steps.}
    \label{tab:result_final}
\end{table*}

\subsubsection{Segmentation}
\label{sec:segmentation_results}
Similar to~\Sref{sssec:segmentation_asr}, we also investigated the impact of audio segmentation for E2E-ST models.
To this end, we used the \texttt{A4} model.
Note that we used the same decoding hyperparameters tuned on Must-C.
The results are shown in Table~\ref{tab:result_st_segmenation}.
We confirmed a similar trend to ASR.
Although $(M_{\rm dur},M_{\rm int})=(1500,100)$ showed the best performance on average, we decided to use $(M_{\rm dur},M_{\rm int})=(2000,100)$ for submission considering the best performance on the latest IWSLT test, tst2019.

\subsubsection{Cascade system}
We also evaluated the cascade system with the Conformer ASR and the Transformer-Base MT trained on the \texttt{WMT10M} data (\texttt{C1}).
The MT model was trained by feeding source sentences without case information and punctuation marks.
The results in Table~\ref{tab:result_st_segmenation} showed that the BLEU scores correlated to the WER in Table~,\ref{tab:result_asr_segmenation} and the performance was comparable with that of \texttt{A4}.
Although there is some room for improving the performance of the cascade system further by using in-domain English LM, it is difficult to conclude which modeling (cascade or E2E) is effective because the cascade system had more model parameters in the ASR decoder and MT encoder.
This means that the E2E model could also be enhanced by using a similar amount of parameters.

\subsubsection{Final system}
Our final system was the best ensemble system \texttt{E4}, using the pyannote.audio based segmentation with $(M_{\rm dur},M_{\rm int})=(2000,200)$\footnote{Because of time limitation, we submitted the systems before completing tuning segmentation hyperparameters.}.
This system, which was our primary submission, scored 24.14 BLEU on tst2019 as shown in Table~\ref{tab:result_final}.
Compared to the result in Table~\ref{tab:result_ensembling}, it improved by 0.53 BLEU thanks to better audio segmentation.
It was also slightly higher than the IWSLT20 winner's submission by SPROL~\citep{potapczyk-przybysz-2020-srpols}.

We also present the results on tst2020 and tst2021 in Table~\ref{tab:result_final}.
Our primary submission \texttt{E4} outperformed the result of last year's winner system on tst2020.

\section{Conclusion}
In this paper, we have presented the ESPnet-ST group's offline systems on the IWSLT 2021 submission.
We significantly improved the baseline Conformer performance with multi-referenced SeqKD, Multi-Decoder architecture, segment merging algorithm, and model ensembling.
Our future work includes scaling training data and careful analysis of the performance gap in different test sets.

\section{Acknowledgement}
This work was partly supported by ASAPP and JHU HLTCOE.
This work used the Extreme Science and Engineering Discovery Environment (XSEDE)~\citep{towns2014xsede}, which is supported by National Science Foundation grant number ACI-1548562.
Specifically, it used the Bridges system~\citep{nystrom2015bridges}, which is supported by NSF award number ACI-1445606, at the Pittsburgh Supercomputing Center (PSC).

\newpage
\bibliographystyle{acl_natbib}
\bibliography{reference}

\end{document}